\documentclass[conference]{IEEEtran}
\IEEEoverridecommandlockouts

\usepackage{cite}
\usepackage{amsmath,amssymb,amsfonts}
\usepackage{algorithmic}
\usepackage{graphicx}
\usepackage{textcomp}
\usepackage{xcolor}
\usepackage{subcaption}
\usepackage{hyperref}
\usepackage{enumitem}
\def\BibTeX{{\rm B\kern-.05em{\sc i\kern-.025em b}\kern-.08em
    T\kern-.1667em\lower.7ex\hbox{E}\kern-.125emX}}
\begin{document}

\title{Harmonic Summation-Based Robust Pitch Estimation in Noisy and Reverberant Environments\\
}

\author{\IEEEauthorblockN{ Anup Singh, Kris Demuynck}
\IEEEauthorblockA{IDLab, Department of Electronics and Information Systems, Ghent University - imec, Belgium \\
\{anup.singh, kris.demuynck\}@ugent.be
}
}

\maketitle

\begin{abstract}
Accurate pitch estimation is essential for numerous speech processing applications, yet it remains challenging in high-distortion environments. This paper proposes a robust pitch estimation method that delivers robust pitch estimates in challenging noise environments. Our approach computes the Normalized Average Magnitude Difference Function (NAMDF), transforms it into a likelihood function, and generates probabilistic pitch states for frames at each sample shift. To enhance noise robustness, we aggregate likelihood values across integer multiples of the pitch period and neighboring frames. Furthermore, we introduce a simple yet effective continuity constraint in the Viterbi algorithm to refine pitch selection among multiple candidates. Experimental results show that our method consistently achieves lower Gross Pitch Error (GPE) and Voicing Decision Error (VDE) across various SNR levels, outperforming existing methods in both noisy and reverberant conditions.
\end{abstract}


\section{Introduction}
Pitch is a fundamental aspect of speech that carries critical linguistic and speaker-specific information. Accurate estimation of the pitch, a.k.a fundamental frequency (F0), is essential for various speech processing applications. Robust pitch tracking enhances speech intelligibility in noisy conditions \cite{buera2008speech}, improves automatic speech recognition (ASR) performance in both tonal and non-tonal languages \cite{singer1992pitch, huang2000pitch}, and facilitates speaker identification in forensic and security applications.

Pitch tracking in real-world environments is inherently challenging due to disruptions in harmonic and temporal structures caused by additive noise and reverberation. While various pitch estimation methods have been developed to provide reliable estimates, their performance degrades significantly in high-distortion conditions, making them unsuitable for noisy and reverberant speech processing tasks. Time-domain approaches, such as the autocorrelation function (ACF) and its variants \cite{article, talkin1995robust}, are particularly susceptible to background noise, frequently resulting in octave errors. While methods like RAPT \cite{talkin1995robust} and its variant implemented in Kaldi \cite{ghahremani2014pitch1} improve voicing decisions, their robustness remains limited in highly degraded environments. Similarly, YIN \cite{de2002yin} refines pitch estimates through post-processing but lacks explicit voiced/unvoiced classification, restricting its effectiveness in practical applications. Frequency-domain methods extract pitch from harmonic peaks \cite{schroeder1968period}, yet they remain vulnerable to octave errors and noise interference. Among these, harmonic summation-based approaches, such as subharmonic summation (SHS) and harmonic product spectrum (HPS) \cite{hermes1988measurement, noll1970pitch}, have demonstrated strong performance in clean and moderately noisy conditions by effectively leveraging harmonic structure for pitch estimation. These methods are particularly advantageous in capturing periodicity information across multiple harmonics, reducing octave errors compared to single-peak detection techniques. SHRP \cite{sun2002pitch} and PEFAC \cite{gonzalez2014pefac} further enhance robustness by refining harmonic summation strategies. However, PEFAC’s reliance on a long analysis window (90.5 ms) limits its applicability for real-time tasks, and SHS-based methods remain sensitive to severe noise and reverberation. While methods like SWIPE \cite{camacho2008sawtooth} and TAPS \cite{huang2012pitch} attempt to improve noise robustness, their dependence on harmonic peak detection makes them less effective in highly degraded environments where harmonics are heavily masked. 
Despite these advancements, existing approaches struggle to maintain accuracy in low-SNR conditions ($leq$10 dB SNR), particularly in the presence of strong reverberation. This limitation underscores the need for a more robust pitch estimation method capable of reliably tracking pitch trajectories even in severe acoustic distortions.

This paper presents a novel monophonic pitch tracking method designed to be robust against high noise and reverberant environments. Our method computes the  Normalized Average Magnitude Difference Function (NAMDF)  as a correlation measure between speech segments. Our method consists of two key components that contribute to the noise robustness. First, we aggregate harmonics evidence for F0 at lags corresponding to integer multiples of the pitch period, a strategy commonly employed in frequency-domain methods to reinforce harmonic structure and improve noise robustness. However, this technique has not been previously explored in time-domain approaches. Second, we transform raw pitch evidence into likelihoods using a sigmoid function, which helps differentiate pitch states corresponding to periodic noise and speech, mitigating the impact of background interference. Additionally, since the harmonic structure of speech evolves more gradually than noise, we further enhance pitch tracking by aggregating pitch evidence across neighboring frames. To address pitch octave errors, we introduce a Viterbi-based decoding algorithm with a straightforward path cost and continuity constraint, ensuring a more reliable pitch trajectory. We evaluate our method on the TUG and Keele databases, demonstrating that our approach significantly outperforms existing baselines in both noisy and reverberant conditions.

\section{Pitch estimation algorithm for speech}
\subsection{Preprocessing} 

We first apply a low-pass filter to the speech signal and segment it into frames using a Hanning window of length \( wF_s \), where \( w \) is the window duration in seconds and \( F_s \) is the sampling frequency. Typically, frames are computed with a stride of \( z = tF_s \) samples, where \( t \) is 5–10 ms. However, our method estimates pitch at every sample, setting \( z = 1 \) to achieve high-resolution pitch tracking and minimize spurious errors. Additionally, each frame is energy-normalized to ensure its peak amplitude falls within \([-1,1]\).

\subsection{Pitch state likelihood} 

\textbf{Computing the likelihood.}
We first determine the lag range for computing the NAMDF based on the frequency search limits and the number of harmonics considered. Let \( F_{\min} \) and \( {F}_{\max} \) denote the minimum and maximum fundamental frequencies, respectively, and let \( H \) be the number of harmonics. The lag range is then defined as:
\begin{equation}
l_{\min} = \frac{F_s}{F_{\max}}, \quad l_{\max} = (H+1) \cdot \frac{F_s}{F_{\min}}
\end{equation}
For each frame \( f_i \), we define the NAMDF for a lag \( l \) as:

\begin{equation}
    \label{eqn:NAMDF}
    \phi^{i}_{l} = \frac{\sum |f_i - f_{i+l}|}{\sqrt[4]{\|f_i\|_2^2  \|f_{i+l}\|_2^2}},  \forall \;l_{min} \leq l \leq l_{max},
\end{equation}

Furthermore, we transform the computed values into likelihoods using the sigmoid function, which helps suppress correlation values arising from periodic noise components. Let $\mathbf{\Phi^i} = (\phi^{i}_{l_{min}}, .., \phi^{i}_{l_{max}})$ represent the NAMDF values for a frame $f_i$. These values are then converted into likelihoods as:
\begin{equation}
    \label{eqn:NAMDF_to_prob}
    \phi^{i}_{l} = \frac{1}{1 + \exp(-(k \frac{\phi^{i}_{l} - \frac{\eta_{.9}+\eta_{.10}}{2}}{\eta_{.90}-\eta_{.10}}) )},\quad \forall \; l_{min} \leq l \leq l_{max}
\end{equation}

where $\eta_{.10}$, $\eta_{.90}$ are 10\textsuperscript{th} and 90\textsuperscript{th} percentile of $\mathbf{\Phi^i}$, respectively and $k$ is a scaling factor. 

The 10\textsuperscript{th} and 90\textsuperscript{th} percentiles are chosen to enhance robustness against outliers and signal variations. The 10\textsuperscript{th} percentile helps suppress noise-induced fluctuations, while the 90\textsuperscript{th} percentile captures strong pitch correlations. By centering the sigmoid function at their mean and scaling it based on their difference, we adaptively distinguish pitch components from noise, ensuring stable and reliable likelihood estimation. 

\textbf{Harmonic Summation.} To enhance pitch estimation, our method aggregates periodicity evidence by incorporating harmonic information at integer multiples of the pitch period as:
\begin{equation}
    \label{eqn:HS}
    \phi^{i}_{l} = \phi^{i}_{l} + \sum_{h=2}^{H+1} w_n  \max(\phi^{i}_{hl \pm r} ), \quad \forall \; l_{min}\leq l \leq l_{max}/(H+1), 
\end{equation}

where $w_n$ are weighting factors that control the contribution of higher-order harmonics. Since harmonic peaks are not strictly aligned with integer multiples of the pitch period, we introduce a tolerance range  $r$ to account for slight variations, ensuring robust pitch detection.

The harmonic summation is performed only for lags where at least \( H \) harmonic likelihood values exist in $\mathbf{\Phi^i}$, while the remaining lags are discarded. This process results in a refined representation $\mathbf{\Phi^i}$, which encapsulates the weighted sum of harmonic likelihoods for each pitch state, spanning the frequency range from \( F_{\min} \) to \( F_{\max} \).

\textbf{Temporal accumulation.} Since pitch variations in speech typically evolve gradually across consecutive frames, we leverage this temporal continuity by accumulating likelihood values over successive frames. We incorporate the temporal accumulation in the pitch likelihood $\phi^{i}_{l}$ as:
\begin{equation}
    \label{eqn:TAPS}
    \phi^{i}_{l} = \sum_{k=-K}^{K} \phi^{i+k}_{l},
\end{equation}
where $K$ defines the window size for temporal integration.

Finally, we obtain the likelihood value corresponding to each pitch state per frame, which encodes the correlation, harmonic strength, and temporal information. 
\subsection{Viterbi  decoding}
In the previous step, we computed the likelihood values for all pitch states within each frame, identifying potential pitch candidates. To generate a smooth and continuous pitch trajectory while eliminating spurious detections, we employ the Viterbi algorithm \cite{viterbi1967error} in the post-processing stage. Since our method computes pitch candidates at a high temporal resolution, we impose a transition constraint that allows only adjacent states in consecutive frames. This constraint ensures smooth pitch variation, preventing abrupt changes and enhancing robustness. Additionally, we apply non-linear upsampling to refine the pitch state resolution and estimate their corresponding likelihood values. Specifically, we interpolate likelihoods at a geometrically increasing sequence of lag values, defined as:

\begin{equation}
    \label{eqn:lags_up}
     l_t = l_{min} (\frac{l_{max}}{l_{min}})^t, \;  0\leq t \leq 1, \; \Delta t=\frac{1}{U*(l_{max}-l_{min})},
\end{equation}
where $U$ is the upsampling factor.

Let $\phi^{i}_t$ represent the likelihood associated with lag $l_t$ for the frame $f_i$. The transition cost between pitch states corresponding to lags $l_t$ and $l_{t'}$ is then defined as:

\begin{equation} \label{eqn:score} C^i_{t,t'} = \phi^{i+1}_{t'} - \phi^{i}_{t}. \end{equation}

Finally, we determine the optimal sequence of pitch states by applying the Viterbi algorithm with the defined cost function.

\subsection{Likelihood Rectificaton}
Since the likelihood values indicate the strength of periodicity within a speech segment, they are inherently unreliable in transition regions, such as shifts from voiced to unvoiced speech and vice versa. Also, likelihood values tend to decrease in regions where the pitch undergoes changes or when pitch jitters are present. This decline in likelihood values can lead to an increased false rejection rate in voicing decisions.

\begin{figure}[t]
    \centering
    \includegraphics[width=0.35\textwidth]{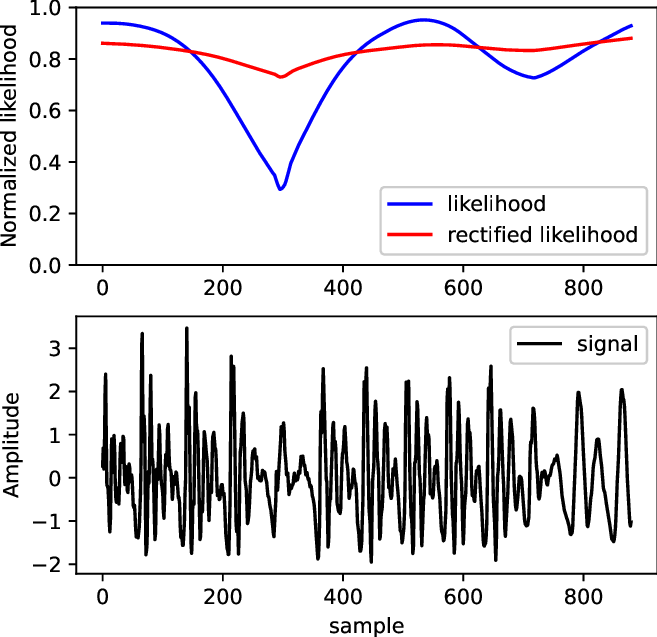}
    \caption{Correction of low-likelihood values (top) caused by irregular periodicity in voiced speech (bottom).}
    \label{rectify}
    
\end{figure}

Figure \ref{rectify} illustrates these challenges by showing the sequence of likelihood values corresponding to the selected pitch states obtained from the Viterbi algorithm. The figure also presents the rectified likelihood values after post-processing, which mitigates these issues. The post-processing procedure is formulated as follows:
\begin{equation}
\begin{split}
    \phi_i & = \phi_i \quad \text{if}\quad c < S, \quad \text{otherwise,}\\ 
    \phi_j & = \alpha \phi_j + (1-\alpha)\phi_{avg} \quad  \forall j = i, ... ,i+J, 
\end{split}
\label{rectified_phi}
\end{equation}

where $c$ represents the number of consecutive frames with likelihood values exceeding a threshold, defined as $t = \max(\mathbf{\Phi})/2$, with $\mathbf{\Phi}$ denoting the likelihood values corresponding to the selected pitch states. The parameter $J$ specifies the number of consecutive future frames over which smoothing is applied. The term $\phi_{avg}$ is the average of the preceding $S$ likelihood values, and $\alpha$ is the weighing factor.

\subsection{Voice Activity Detection}
The likelihood values corresponding to the decoded pitch states are used to classify each frame as voiced or unvoiced. However, we observe that the likelihood values remain relatively high in silent regions, particularly due to the presence of periodic noise components at very low SNR levels. To this end, we introduce a voicing factor that attenuates the likelihood values in unvoiced regions. This factor is derived using two acoustic features: frame energy and an NAMDF-based feature, computed as follows:
\begin{equation}
    \label{eqn:cNAMDF}
   \omega_i = log\bigg(\max_{j}\sum_{l=j}^{j+W-1} \phi^{i}_{l}\bigg),
\end{equation}
where $\phi^{i}_{l}$ is the raw likelihood value obtained in Eq. \ref{eqn:TAPS}.

The two-element feature vectors extracted for each frame are first transformed into a single feature using PCA \cite{jolliffe2016principal}. A bimodal Gaussian Mixture Model (GMM) is then fitted to the transformed features to distinguish between voiced and unvoiced regions. The voicing factor for a frame $f_i$ is computed as: 
\begin{equation}
    \label{eqn:voice_fac}
    v_i = {\bigg(1 + \frac{p_{i,1}}{p_{i,2}}\bigg)}^{-1}
\end{equation}
 where $p_{i,1}$ and $p_{i,2}$ represent the likelihoods corresponding to voiced and unvoiced components of the GMM, respectively. Finally, the rectified likelihood value $\phi_i$ from Equation \ref{rectified_phi} is scaled down using the voicing factor as $V_i = \phi_i * v_i$. Finally, this value is normalized to lie within the range [0,1], providing the estimated voicing probability.

\section{Implementation Details}
\subsection{Databases}
\begin{itemize}[leftmargin=*]
    \item \textbf{TUG Database}  \cite{pirker2011pitch}: This dataset comprises speech signals sampled at 48 kHz from 20 native English speakers (10 male, 10 female). It includes 4,720 recorded sentences, totaling 9.6 hours of speech. The reference F0 estimates are provided at 10 ms intervals, extracted from corresponding laryngograph signals using the RAPT method \cite{garofolo1993darpa}.

    \item \textbf{Keele Database} \cite{plante1995pitch}: This dataset contains recordings from 10 male and 10 female speakers at a 20 kHz sampling rate, with a total duration of 9 minutes \cite{plante1995pitch}. Reference F0 estimates, extracted from parallel laryngograph signals at 10 ms intervals, exhibit inaccuracies, particularly in transient regions. Despite these limitations, we include the Keele database for comparability with studies that have used it.

    \item \textbf{Noises}: We utilize the ETSI database \cite{etsi} for noise signals, in which we choose seven real-life background noise types: \textit{Babble}, \textit{Living Room}, \textit{Cafeteria}, \textit{Car}, \textit{Workplace}, \textit{Traffic}, \textit{Train Station}. Additionally, we include \textit{White Gaussian} noise as a stationary noise source.
\end{itemize}

\subsection{Metrics}

\begin{itemize}[leftmargin=*]
    \item \textbf{Gross Pitch Error (GPE)} \cite{hu2010tandem}: provides the accuracy measure of pitch estimate in the voiced region. 
    \begin{equation}
        GPE = \frac{N_{0.05}}{N_{v}}
    \end{equation}
    where $N_{0.05}$ is the number of frames with pitch estimates that differs by more than 5\% from ground truth, and $N_{v}$ is the total number of voiced ground truth
frames. 

    \item \textbf{Voicing Decision Error} (VDE) \cite{fisher2006generalized}: indicates the voicing detection performance. It is computed as the percentage of total frames misclassified in terms of voicing.  
    \begin{equation}
        VDE = \frac{N_{ve} + N_{uve}} {N} 
    \end{equation}
    where $N_{ve}$ and $N_{uve}$ denote the number of frames misclassified as voiced and unvoiced frames, respectively, and $N$ is the total number of ground truth frames.
\end{itemize}

\subsection{Baselines}
We select both time-domain and frequency-domain pitch extraction baselines, including PEFAC, YIN, SHRP, SWIPE, Kaldi, and the Autocorrelation (AC) and Cross-correlation (CC) methods from Praat. Each baseline is evaluated using its default parameter settings. Notably, Praat, SHRP, and SWIPE assign a pitch value of zero to frames classified as unvoiced. To ensure non-zero pitch estimates for every frame—necessary for computing the GPE rates—we disable their respective voicing detection modules.

\section{Results and Discussion}

\begin{figure}[t]
    \centering
    \begin{subfigure}{0.49\columnwidth}
        \centering
        \includegraphics[width=\linewidth]{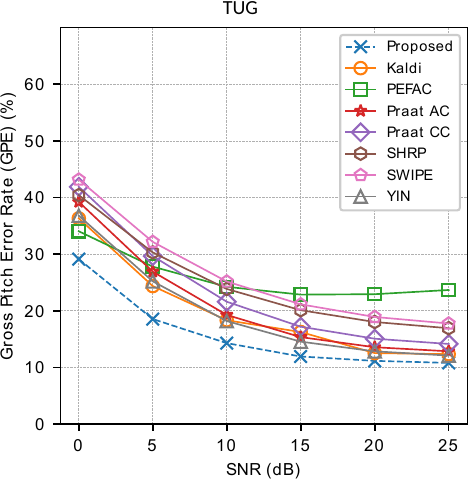}
        \label{fig:sub1}
    \end{subfigure}
    \hfill
    \begin{subfigure}{0.49\columnwidth}
        \centering
        \includegraphics[width=\linewidth]{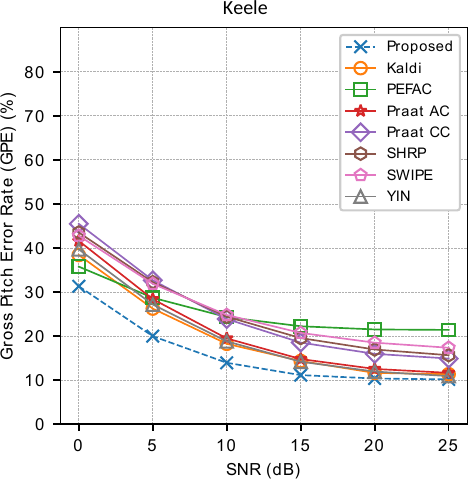}
        \label{fig:sub2}
    \end{subfigure}
    \caption{Gross Pitch Error (GPE) rates across different noise levels for the TUG and Keele databases.}
    \label{fig:GPE_combined}
\end{figure}

Figure \ref{fig:GPE_combined} compares our method with baseline approaches in terms of GPE, averaged across different noise types and SNR levels. Our method consistently achieves the lowest GPE, with the most significant improvement observed in low-SNR conditions ($\leq$10 dB). At these noise levels, our approach reduces GPE by up to 15\% compared to the best-performing baseline. As SNR increases, the performance gap narrows, but our method maintains an advantage.

Among different noise types, car noise resulted in the highest errors at low SNRs. However, our method demonstrates strong robustness to car noise, achieving up to 30\% lower GPE than baselines in these conditions. Additionally, baseline methods frequently suffer from pitch-halving errors in female utterances due to dominant low-frequency noise. While PEFAC effectively suppresses low-frequency noise and performs well for female speakers, it tends to introduce pitch-doubling errors. In contrast, our method significantly reduces pitch octave errors by leveraging harmonic evidence and enforcing smooth transitions through Viterbi decoding.

We also evaluate the performance of the proposed method in noisy reverberant environments. To simulate these conditions, both speech and noise signals are first filtered using a room impulse response (RIR) with a $t_{60}$ of 0.7s, followed by the addition of noise at varying SNR levels. Tables 1 and 2 present the GPE results for the TUG and Keele databases, respectively. As expected, the introduction of reverberation, in addition to noise, significantly increases pitch estimation errors across all methods. However, our approach demonstrates greater resilience, exhibiting less performance degradation compared to the baselines. The relative performance trends observed in noisy environments remain consistent in noisy reverberant conditions. Notably, our method maintains its advantage over most baselines, though it slightly underperforms compared to PEFAC at 0 dB SNR.

\begin{table}[]
    \renewcommand{\arraystretch}{1}
    \centering
     \caption{GPE rates in noisy reverberant conditions for the TUG database}
    \begin{tabular}{c|c|c|c|c|c|c}
        \hline
         SNR (dB) & 0 & 5 & 10 & 15 & 20 & 25 \\
        \hline
        \hline
         Proposed method & 0.57 & \textbf{0.42} & \textbf{0.34} & \textbf{0.32} & \textbf{0.31} & \textbf{0.31} \\
         Kaldi & 0.68 &  0.54 & 0.43 & 0.37 & 0.35 & 0.35 \\
         YIN & 0.61 & 0.47 & 0.38 & 0.36 & 0.33 & 0.33\\
         PEFAC & \textbf{0.55} & 0.48 & 0.46 & 0.45 & 0.45 & 0.46 \\
         Praat (AC) & 0.66 & 0.54 & 0.44 & 0.38 & 0.37 & 0.36\\
         Praat (CC) & 0.67 & 0.55 & 0.46 & 0.39 & 0.38 & 0.37\\
         SHRP & 0.60 & 0.51 & 0.46 & 0.44 & 0.43 & 0.43\\
         SWIPE & 0.58 & 0.48 & 0.42 & 0.39 & 0.38 & 0.37\\
         \hline
    \end{tabular}
   
    \label{tab:nosise_rev_tug}
\end{table}

\begin{table}[]
    \renewcommand{\arraystretch}{1}
    \centering
    \caption{GPE rates in noisy reverberant conditions for the Keele database}
    \begin{tabular}{c|c|c|c|c|c|c}
        \hline
         SNR (dB) & 0 & 5 & 10 & 15 & 20 & 25 \\
        \hline
        \hline
         Proposed method & 0.54 & \textbf{0.41} & \textbf{0.33} & \textbf{0.31} & \textbf{0.31} & \textbf{0.30} \\
         Kaldi & 0.60 &  0.45 & 0.36 & 0.32 & 0.31 & 0.31 \\
         YIN & 0.58 & 0.46 & 0.37 & 0.33 & 0.33 & 0.32\\
         PEFAC & \textbf{0.51} & 0.44 & 0.41 & 0.40 & 0.39 & 0.39 \\
         Praat (AC) & 0.64 & 0.51 & 0.41 & 0.37 & 0.35 & 0.35\\
         Praat (CC) & 0.66 & 0.53 & 0.43 & 0.40 & 0.38 & 0.38\\
         SHRP & 0.59 & 0.50 & 0.45 & 0.43 & 0.42 & 0.42\\
         SWIPE & 0.55 & 0.46 & 0.41 & 0.38 & 0.37 & 0.36\\
         \hline
    \end{tabular}
    
    \label{tab:noise_rev_keele}
\end{table}

\begin{figure}[t]
    \centering
    \begin{subfigure}{0.49\columnwidth}
        \centering
        \includegraphics[width=\linewidth]{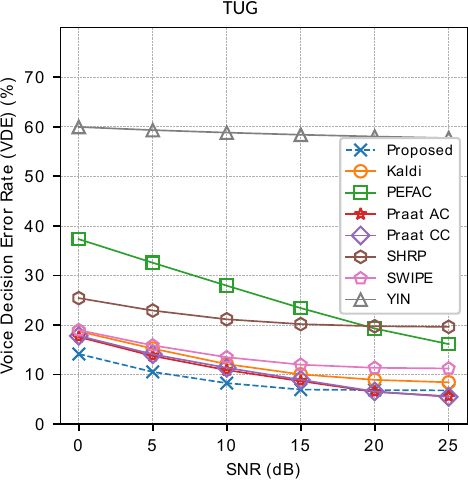}
        \label{fig:sub1}
    \end{subfigure}
    \hfill
    \begin{subfigure}{0.49\columnwidth}
        \centering
        \includegraphics[width=\linewidth]{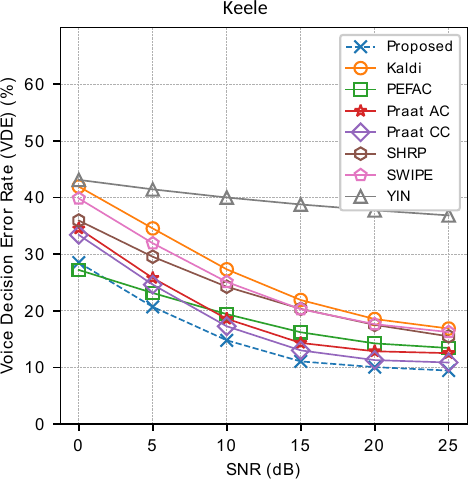}
        \label{fig:sub2}
    \end{subfigure}
    \caption{Voicing Decision Error (VDE) rates across different noise levels for the TUG and Keele databases.}
    \label{fig:VDE_combined}
\end{figure}

Figure \ref{fig:VDE_combined} presents the voicing detection performance averaged across different noise types and SNR levels for both databases. Our method consistently outperforms the baselines, particularly in high-noise conditions (SNR $\leq$ 10 dB), where the performance gap is most pronounced. As the SNR increases, the gap between our method and the baselines narrows, indicating that all methods perform better in cleaner conditions. These results demonstrate that our voicing detection approach, despite its simplicity, is highly effective in handling noisy environments,

\begin{figure}
    \centering
    \includegraphics[width = 0.3\textwidth]{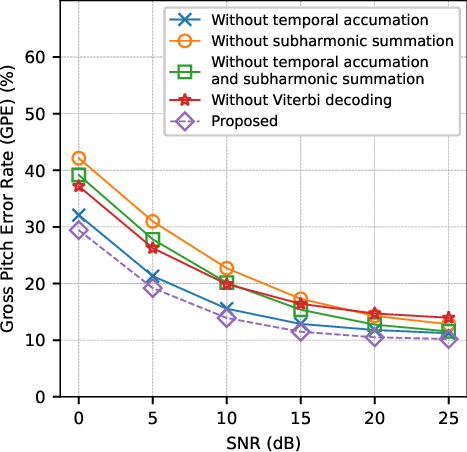}
    \caption{Impact of each proposed component on pitch estimation performance.}
    \label{breakdown_analysis}
\end{figure}

Figure \ref{breakdown_analysis} illustrates the impact of each proposed component in our pitch estimation method in terms of GPE(\%). In the absence of the Viterbi algorithm in the post-processing stage, the pitch state with the highest likelihood is selected as the estimated pitch. The results clearly show that harmonic summation significantly enhances performance, reducing the GPE by up to 13\%. Temporal accumulation further improves performance, yielding a ~2\% GPE reduction up to 10 dB SNR, with diminishing returns at higher SNRs. Notably, removing both components leads to a substantial performance drop, highlighting their combined effectiveness in noise-robust pitch estimation. Also, incorporating the Viterbi algorithm in the post-processing stage further refines pitch trajectories, demonstrating its importance in stabilizing pitch estimates.

Due to space limitations, additional supplemental information is available on the webpage.\footnote{\url{https://github.com/anupsingh15/HSRPE}}

\section{Conclusion}

In this paper, we proposed a robust time-domain-based F0 detection method that delivers reliable pitch and voicing estimates, even in challenging noisy conditions. Our approach leverages the Normalized Average Magnitude Difference Function (NAMDF) as a correlation measure between speech segments, which is then transformed into a likelihood function using a sigmoid mapping to mitigate spurious pitch estimation errors. To enhance robustness against noise and reverberation, we incorporate harmonic structure and temporal continuity constraints. We employ the Viterbi algorithm with added constraints in the post-processing stage to obtain a smooth pitch trajectory. Our method also estimates voicing likelihood based on NAMDF and frame energy features. Experimental results demonstrate that our approach consistently outperforms baseline methods, particularly in low-SNR conditions ($\leq$10 dB) in noisy as well as noisy reverberant environments.



\bibliographystyle{IEEEtran}


\end{document}